# Documentation of Machine Learning Software

Yalda Hashemi*, Maleknaz Nayebi*, Giuliano Antoniol*
*Dept. computer and software engineering
Ecole Polytechnique de Montreal, Canada

{yalda.hashemi, maleknaz.nayebi, giuliano.antoniol}@polymtl.ca

*Abstract*—Machine Learning software documentation is different from most of the documentations that were studied in software engineering research. Often, the users of these documentations are not software experts. The increasing interest in using data science and in particular, machine learning in different fields attracted scientists and engineers with various levels of knowledge about programming and software engineering. Our ultimate goal is automated generation and adaptation of machine learning software documents for users with different levels of expertise. We are interested in understanding the nature and triggers of the problems and the impact of the users' levels of expertise in the process of documentation evolution. We will investigate the Stack Overflow Q&As and classify the documentation related Q/As within the machine learning domain to understand the types and triggers of the problems as well as the potential change requests to the documentation. We intend to use the results for building on top of the state of the art techniques for automatic documentation generation and extending on the adoption, summarization, and explanation of software functionalities.

*Index Terms*—Software Engineering, Machine Learning, Software Documentation, Mining Software Repositories.

## I. INTRODUCTION

Documentation is meant to guide users about software functionality. Documentation is "any artifact which its purpose is to communicate information about the software system to which it belongs, to individuals involved in the production of that software" [2]. State of the art of software engineering discussed methods for automated generation of software documentation, for example, API descriptions.

The increasing growth in the use of machine learning (ML) in a variety of domains requires data scientists to familiarize themselves with the code and software structure. This says, for proper use of ML techniques, a data scientist in biomedical or climate engineering should be able to ideally refer to the software documentation to understand how to use the product (for example, an API in TensorFlow). However, the level of technicality in current software documentation urges the users to look for additional resources continuously. These questions often find their way to StackOverflow (SO).

Our research is intended to provide the users of ML software with clear and thorough information about the software product in a way being understandable in consideration of their level of expertise [5]. We are intent to ultimately adopt state of the art methods for automated construction of software documentation to develop *expertise-aware software documentation* for ML products. We answer three research questions:

**RQ1:** What type of problems do users face when using documentation of ML software?
**RQ2:** How does the expertise level of users impact their understandings of the documentation of ML software?
**RQ3:** How does documentation evolve in relation to the SO questions?

## II. METHODOLOGY

### A. RQ1: Nature and triggers of ML documentation problems

SO Q/As has been a common source for researchers to understand developers' problems and pain points, in general, and in particular with ML software tools [1], [3]. Being interested in understanding the current usefulness of documentation related to ML software, we mine the SO questions to identify:

**What** types of problems do users have with ML software, and in what areas?
**Which** types of documentations are used and questioned?
**Why** the documentations are referred to, and in which area?
**How** the documentation referrals happen?

Figure 1 shows a sample question from SO and the annotation showing what the problem is, which documentation is used and how it has been referred to. To systematically answer these questions, we start by gathering the SO Q/As tagged as ML or popular ML libraries such as Tensorflow or Pytorch. We then manually annotate a representative sample of this data as being related or unrelated to the software documentation, using at least two annotators. Having the questions identified as documentation related, we perform thematic analysis to identify what, why, how, and which questions by manual analysis of the categories using combination of open and closed card sorting. The results would provide a taxonomy of problems and their attributes. To answer the "What" question with regards to the ML problems, we use the categories introduced in [3].

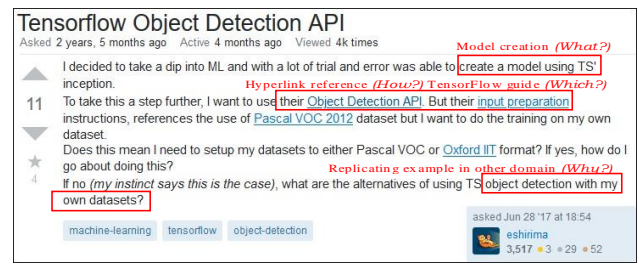

Fig. 1. Example question and the information being analyzed in this study

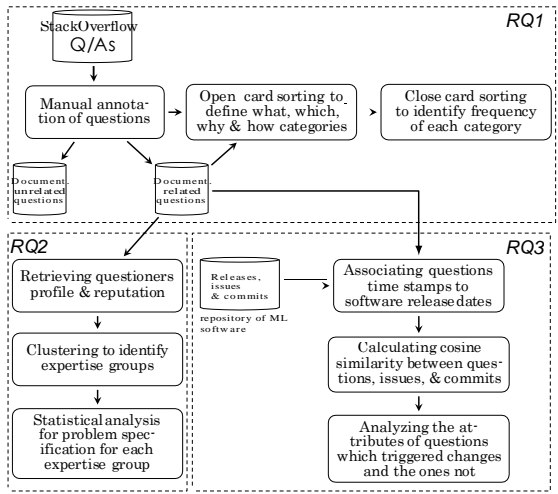

Fig. 2. Methodology for analyzing documentation evolution in ML software

*B. RQ2: the impact of level of expertise of users on their understanding from ML software documentation*

We are interested in evaluating if developers understanding of software and the domain (here ML) impacts their question types. Learning this, we intend to ultimately adapt the software documentation for different user groups and their information needs. To this aim, we use the factors introduced by Movashovitz et al. [4] (the number of up voted questions, number of up voted answers, number of accepted answers, number of down voted answers). To identify users level of expertise in ML, we also consider the number of posts tagged in relation to ML (for example, tensorflow or machinelearning). Having this, we perform clustering using k-means (identifying $k$ with the elbow method). As for the last stage, we perform statistical test to compare groups with different levels of expertise and reason out the extent of difference between their needs from software documentation. Figure 2 shows the main steps of this process.

*C. RQ3: Co-evolution of documentation and questions*

To understand if and to what extent the raised questions triggered a change in the documentation of ML software, we connect the questions to the opened issues in the respective software repository and the commit messages while changing the documentation[1]. In this RQ, we are interested to see if a question from SO has triggered a change in the documentation or not and if so what is the characteristics of these types of questions. The process is shown in Figure 2. First, we associate each question to a release of a software assuming that a valid question that is asked before *Release$_i$* can not be responded in *Release$_j$*, where $j < i$. After, we calculate the cosine similarity between the text of the SO question and the issues in the repository. For those questions not matching any issues, we analyze the similarity between the commit messages and the questions. Tracing a question to the documentation changes, we look into the question attributes in RQ1 and RQ2 to understand the co-evolution of users understandings and documentations.

[1]ML software often have document repository for example https://github.com/tensorflow/docs.

### III. SUMMARY AND FUTURE WORK

We intend to automate the generation of software documentation in the ML domain in consideration of different levels of user expertise. To achieve that, we take initial steps by understanding users behavior in using ML documentation and understanding the process of evolution of the software documents in ML domain.

So far, we took steps toward answering RQ1 and performed a preliminary study on "TensorFlow" as one of the popular ML libraries [3] (TensorFlow has 48,122 questions on SO). By randomly sampling 500 questions with the tensorflow tag and manually categorizing the questions, we found 16.6% of these questions are related to documentation. We performed a two annotator light-weighted card sorting process to check our hypothesis of RQ1. We observed that SO questions are mainly concerned with parameter tuning (22.6%), model creation (14.2%), and error/exception (10.7%) when categorized along with ML domain as defined by Islam et al. [3] *(what)*. Most of the questions have been triggered as the users were not able to replicate the examples provided in the documentation (24.8%) and 12.9% were concerned with the lack of description on the implementation and use of the software in documentation *(why)*. Also, we saw 60.7% of these documentations are related to the official TensorFlow documentation, while others refer to third party material such as tutorials, videos, books and scientific papers *(which)*. We also observed that the majority of these questions (72.8%) hyperlinked to the mentioned documentation, while several others used a screenshot or just mentioned the name of the documentation. We extend the study by enlarging the sample size and across other popular libraries, including "TensorFlow" and "Pytorch". The result of this study will provide us with the knowledge to develop, simplify, and summarize the documentation for a better understanding of its users and later developing an automated tool for *expertise-aware software documentation* for ML products.